\documentclass[sigconf]{acmart}
\AtBeginDocument{%
  }

\copyrightyear{2026}
\acmYear{2026}
\setcopyright{cc}
\setcctype{by}
\acmConference[CIKM '26]{Proceedings of the 35th ACM International Conference on Information and Knowledge Management}{November 7--11, 2026}{Rome, Italy.}
\acmBooktitle{Proceedings of the 35th ACM International Conference on Information and Knowledge Management (CIKM '26), November 7--11, 2026, Rome, Italy}
\acmISBN{979-8-4007-2539-5/2026/11}
\acmDOI{10.1145/3799682.3840135}
\usepackage{xspace}
\usepackage{multirow}
\usepackage{graphicx}
\usepackage{subcaption}
\usepackage{cleveref}
\usepackage{enumitem}
\usepackage{colortbl}
\usepackage[ruled,vlined]{algorithm2e}
\usepackage{makecell} 

\begin{document}

\title{Pre-retrieval Query Clustering for Adaptive Top-k Document Retrieval in RAG Systems}

\author{Ye Xia}
\authornote{This work was performed when the author was at EvenUp.}
\orcid{0009-0003-3817-6087}
\affiliation{%
  \institution{Portland State University}
  \city{Portland}
  \state{OR}
  \country{USA}
}
\email{xia2@pdx.edu}

\author{Emre Yamangil}
\orcid{0000-0001-5606-233X}
\affiliation{%
  \institution{EvenUp}
  \city{San Francisco}
  \state{CA}
  \country{USA}
}
\email{emre.yamangil@evenup.ai}

\author{Haixun Wang}
\orcid{0000-0002-1378-4241}
\affiliation{%
  \institution{EvenUp}
  \city{San Francisco}
  \state{CA}
  \country{USA}
}
\email{haixun.wang@evenup.ai}
\renewcommand{\shortauthors}{Ye Xia, Emre Yamangil, \& Haixun Wang}

\begin{abstract}
RAG systems commonly retrieve a fixed number of documents (top-k) to ground generation, but this static approach is brittle: simple queries suffer over-retrieval (adding noise and cost) while complex queries are under-retrieved, causing recall failures that cascade into incorrect answers.  Motivated by the question of how many documents must be retrieved to answer an arbitrary query reliably, we propose a practical, general framework for query-adaptive retrieval depth. Offline, we estimate per-query retrieval difficulty by measuring NDCG under the default retriever and deriving a query-specific “saturation” point k* from the NDCG–k curve. Because computing these signals online is expensive, we cluster a large set of queries in embedding space and summarize each cluster with a recommended retrieval depth that targets high coverage (e.g., \textasciitilde{}95\%) using a mean-plus-variance rule. At runtime, the system assigns an incoming query to a cluster and selects the corresponding top-k in constant time. Compared with post-retrieval confidence methods that rely on clustering retrieved documents, our approach is pre-retrieval and query-centric, making it robust in heterogeneous, case-like corpora and applicable across domains such as legal, healthcare, finance, and enterprise search. Finally, this framework has been tested in full-traffic queries that improved $F_1$ by over 36\% while reducing token usage by 14\% on low-complexity clusters without accuracy loss.
\end{abstract}



\maketitle

\section{Introduction}
\label{sec:Introduction}
Adaptive retrieval depth is a general-purpose problem in Retrieval-Augmented Generation (RAG) systems. RAG grounds LLM responses in external knowledge---documents, manuals, code, internal wikis---by retrieving relevant material and placing it into the model's context before generation. In practice, most RAG systems still rely on a \emph{static top-$k$} retrieval rule \cite{gao2024retrieval, lewis2020retrieval}: encode the query, retrieve the $k$ nearest documents, and feed them all to the model as context. While simple and robust, static top-$k$ is fundamentally mismatched to real user behavior: \emph{narrow questions often need only a small, precise snippet}, while broad, ambiguous, or multi-part questions may require many pieces of evidence.

Using legal tech as a concrete example, the same RAG assistant can receive queries whose ``right'' $k$ differs by an order of magnitude. A system might receive a \emph{simple lookup} like:
\begin{itemize}
  \item \textit{``What is the statute of limitations for personal injury in California?''} (often answered by one authoritative source)
\end{itemize}
or a \emph{moderately contextual} question like:
\begin{itemize}
  \item \textit{``Explain how comparative negligence affects damages when the plaintiff is partially at fault, and cite the relevant rule.''} (may require multiple sources: doctrine + jurisdiction-specific nuance + definitions)
\end{itemize}
or a \emph{complex synthesis} request like:
\begin{itemize}
  \item \textit{``Given these facts (timeline, injuries, treatment, insurance limits), draft a demand summary and explain what evidence supports each damages category.''} (may require many documents: medical records, billing, prior demands, policy language, deposition excerpts, relevant jury instructions, etc.)
\end{itemize}

Static top-$k$ fails in two symmetric ways. If $k$ is too small, the system misses critical documents, producing incomplete or incorrect responses. If $k$ is too large, context becomes bloated with redundant or weakly related information, increasing token cost and latency while potentially degrading answer quality. \emph{Retrieval depth should vary by query}.

How many documents should we retrieve for a given query---enough to be complete, but not so many that we drown the model in noise? One representative post-retrieval approach is \emph{Cluster-based Adaptive Retrieval (CAR)} \cite{xu2025clusterbasedadaptiveretrievaldynamic}. CAR starts from a large candidate list and  selects an optimal cutoff by analyzing clustering patterns in the ordered query--document similarity distances. Intuitively, CAR looks for transition points in the ranked similarity curve---where a tight group of highly relevant documents gives way to less pertinent candidates---and sets cutoffs aligned to those breakpoints. It clusters the ranked distance values and scores cluster boundaries using gap-based metrics plus position penalties to avoid premature truncation. In a deployed Coinbase CDP setting, CAR initializes with 40 candidates and adaptively filters to roughly 16 documents on average, yielding substantial reductions in token usage, latency, and hallucination rate while preserving response quality.

However, post-retrieval approaches like CAR inherit a key operational constraint: they must retrieve a relatively large candidate pool first, then decide what to keep. This can be suboptimal when (i) retrieval itself is expensive (multi-stage retrieval, cross-encoders, heavy filtering), (ii) corpora contain many near-duplicates or repetitive templates, or (iii) systems need predictable budgets and latencies before hitting downstream components. Moreover, post-retrieval signals depend on the geometry of retrieved results, which varies with embedding model choice and corpus structure.

In this work, we propose a general, \emph{query-centric} approach for adaptive top-$k$ control that estimates retrieval depth needs from the \emph{query} (and optionally lightweight metadata) rather than from clustering the retrieved set at inference time. The core idea is to learn \emph{query complexity / required evidence breadth} from historical interactions and offline evaluation, then use it to choose $k$ \emph{before} executing retrieval. Query-centric adaptation is feasible in many real deployments because they operate in a \emph{fixed use case} with relatively bounded task distributions. Consider personal injury law: though each case has unique facts, the \emph{questions users ask} often draw from recurring intent families (e.g., identify key dates, summarize treatment chronology, find liability facts, extract wage loss evidence, support damages arguments with citations). In such settings, we can analyze questions \emph{in advance}: collect historical queries, embed and cluster them by intent, and validate clusters through lightweight review or periodic curation. Once clusters are understood, we attach operational policies to each---including retrieval depth budgets, source preferences, and structured evidence requirements. This shifts the hardest work offline and reduces online adaptation to a fast ``assign query to cluster and apply policy'' step.

Our approach complements post-retrieval methods like CAR: CAR provides a ``posterior correction'' based on observed distance curves; we provide a ``prior'' that sets an initial $k$ early, improving predictability and reducing unnecessary retrieval on simple queries while expanding for complex ones. The two can be combined: pre-retrieval $k$ selection sets an initial budget, followed by CAR-style post-retrieval trimming when beneficial.

In summary, this paper makes the following contributions:
\begin{itemize}
  \item We formalize adaptive top-$k$ selection as a query-centric retrieval depth control problem for RAG across domains (e.g., legal, healthcare, enterprise search).
  \item We introduce a practical method to estimate query complexity offline and map it to retrieval depth at runtime using UMAP clustering and NDCG-driven depth estimation.
  \item We validate the approach empirically: query-adaptive retrieval improves $F_1$ by 36.64\% in an online A/B test, while enabling substantial token reductions (up to 14\%) on low-complexity clusters without accuracy loss and improving evidence coverage (e.g., up to $\sim$17\% more identified treatments) on difficult clusters.
\end{itemize}

The rest of the paper is organized as follows. \Cref{sec:RelatedWork} provides necessary background and discusses related work in information retrieval. \Cref{sec:qcar} presents the design and architecture of the proposed Query-Cluster Adaptive Retrieval framework. \Cref{sec:methodology} details the construction of the Query-Cluster database $\mathcal{D}_{qc}$ and online inference procedure. \Cref{sec:experiments} evaluates performance and presents empirical results. Finally, \Cref{sec:conclusion} concludes and outlines directions for future work.
\section{Related Work}
\label{sec:RelatedWork}
Traditional Information Retrieval (IR) has transitioned from keyword-based models like BM25 \cite{Robertson1994Okapi} to neural dense retrieval models \cite{lewis2020retrieval}, which capture semantic nuances essential for modern RAG. However, most RAG implementations utilize a static top-$k$ approach, which ignores the inherent variance in query complexity, thereby affecting the efficiency and accuracy of the system.

Query Performance Prediction (QPP) seeks to estimate retrieval quality either before or after search retrieval. Early neural pre-retrieval QPP models used geometric properties of embeddings \cite{Arabzadeh2020NeuralEM} or supervised fine-tuning models, such as BERT-QPP \cite{10.1145/3459637.3482063} and BERT-PE ~\cite{Khodabakhsh2024BertPEAB}, to predict if a query would yield relevant results. While these methods identify hard queries, they do not typically prescribe an optimal top-$k$ for downstream tasks. 

The "Cluster Hypothesis" in IR suggests that relevant documents tend to be more similar to each other than to non-relevant ones. This has motivated post-retrieval methods such as using NQC metric ~\cite{ganguly2023queryspecificvariabledepthpooling}, ClusQPP \cite{Akaberi2025ClusQPP}, and CAR \cite{xu2025clusterbasedadaptiveretrievaldynamic}. These models utilize the cohesion and distributions of the retrieved data to perform dynamic cutoffs. While effective, these document-centric methods require an initial large-scale retrieval (the "posterior") before adaptation can occur. As noted in \Cref{sec:Introduction}, this can be computationally expensive with latency issues.

Query-centric adaptation offers a proactive alternative. Frameworks such as HyPA-RAG \cite{kalra2024hypa} classify query complexity to adjust RAG parameters, yet they typically rely on heuristic categories. This leaves a critical gap in the literature: a method that combines the runtime efficiency of pre-retrieval prediction with the precision of empirical calibration.

To bridge this gap, we introduce the \textbf{Query-Clustering Adaptive Retrieval (QCAR)} framework that maps semantic intent to optimal retrieval depths, coupling comprehensive offline database with lightweight online inference. 

\section{Query-Clustering Adaptive Retrieval Framework (QCAR)}
\label{sec:qcar}
\begin{figure*}
\centering
\includegraphics[width=\textwidth]{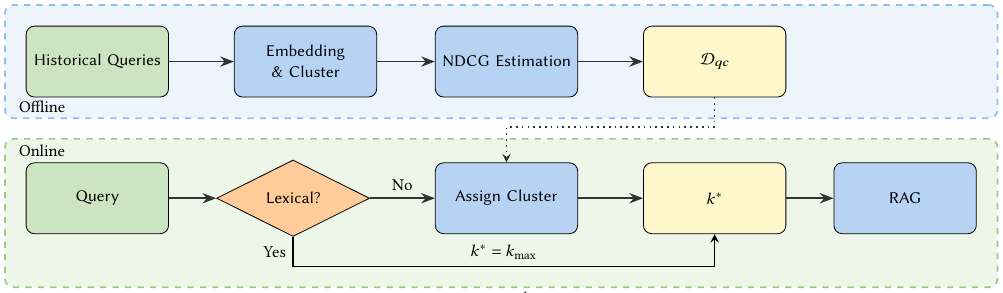}
\caption{QCAR RAG Framework}
 \Description{Diagram of the proposed QCAR framework, showing the offline query-clustering process and the online adaptive retrieval process.}
\label{fig:QCAR}
\end{figure*}

This section presents the conceptual foundation of the Query-Clustering Adaptive Retrieval (QCAR) framework. We begin with the core insight motivating our approach, then characterize the two dimensions along which query complexity manifests, describe how QCAR operationalizes these insights, and finally position our method relative to post-retrieval alternatives.

\subsection{Core Insight: Query Complexity as a Pre-Retrieval Property}
\label{sec:core-insight}

In bounded domains, retrieval depth is largely predictable from the query. The same question type consistently requires similar retrieval depth because corpus structure and relevance distribution are stable. This predictability is the foundation of QCAR.

Consider two queries in a personal injury law system:
\begin{itemize}
    \item \textbf{Easy Queries.} Low retrieval depth, as the target information is highly localized.\\
    \emph{``What is the plaintiff's date of birth?''} --- narrow, fact-seeking, typically satisfied by one authoritative source.
    \item \textbf{Hard Queries.} High retrieval depth due to the sheer breadth of evidence needed to comprehensively satisfy the topic. \\
    \emph{``List all treatments with dates and providers.''} --- broad, exhaustive, requiring comprehensive coverage of the medical record.
\end{itemize}

\noindent These queries have characteristic retrieval depths that generalize across cases because the corpus has predictable structure and query types recur.

In open-domain settings, this predictability weakens. The same query might require vastly different retrieval depths depending on corpus quality, document redundancy, and relevance density. Post-retrieval methods like CAR~\cite{xu2025clusterbasedadaptiveretrievaldynamic}, which analyze the retrieved set to determine cutoffs, are better suited to such settings.

QCAR targets the bounded-domain case, where we can estimate complexity \emph{before} retrieval, enabling predictable latency budgets, early termination for simple queries, and reducing computation.

\subsection{Two Dimensions of Query Complexity}
\label{sec:complexity-dimensions}

What makes a query complex from a retrieval perspective? We identify two complementary signals that jointly determine required retrieval depth. QCAR integrates these through a simple priority rule: \emph{lexical markers act as deterministic overrides; semantic intent provides statistical priors.}

\subsubsection{Lexical Scope Markers}
\label{sec:lexical-markers}

Some queries explicitly encode their retrieval requirements through language: quantifiers (``all,'' ``every,'' ``complete''), cardinality parameters (``top 5,'' ``list 10''), structural iteration (``for each provider,'' ``grouped by''), and synthesis tasks (``timeline,'' ``chronological order,'' ``cite evidence''). These explicit signals reliably indicate high retrieval depth regardless of topic. A query containing ``list all'' requires comprehensive retrieval whether it concerns medical treatments, billing records, or witness statements.

Lexical markers account for approximately 20\% of queries in our dataset. Though a minority, they represent high-stakes cases: missing evidence on a ``list all'' query constitutes a recall failure by definition.

\subsubsection{Semantic Intent}
\label{sec:semantic-intent}

The majority of queries lack explicit scope markers yet vary substantially in retrieval complexity based on \emph{what} they ask. Consider:
\begin{itemize}
    \item \emph{``Who was the driver at fault?''} --- typically answered by a single police report or witness statement.
    \item \emph{``Explain how comparative negligence affects damages in this case.''} --- may require doctrine, jurisdiction-specific rules, case facts, and prior rulings.
\end{itemize}

\noindent Both queries are lexically unmarked yet differ substantially in evidence breadth. This implicit complexity arises from semantic structure and must be learned from domain data.

When queries contain explicit scope markers, we bypass learned estimates and retrieve maximally ($k^* = k_{\max}$). Otherwise, we rely on learned cluster-level recommendations derived from semantic similarity to historical queries.

\subsection{The QCAR Approach: Offline Learning, Online Inference}
\label{sec:qcar-approach}

QCAR decouples the problem into a computationally intensive offline phase and a lightweight online phase.

\subsubsection{Offline: Learning the Query Space}
\label{sec:offline}

In bounded domains, user queries cluster into recurring \emph{intent families}. We exploit this structure through three offline steps:

\begin{enumerate}
    \item \textbf{Discover intent clusters.} Embed historical queries using a dense encoder and cluster by semantic similarity. Queries within a cluster share similar retrieval requirements.
    
    \item \textbf{Measure retrieval difficulty.} For a sample of queries from each cluster, evaluate retrieval quality using NDCG. This quantifies how many documents the retriever must fetch to achieve acceptable recall.
    
    \item \textbf{Assign cluster-level depth recommendations.} Aggregate per-query difficulty estimates into a single recommended $k^*$ for each cluster, targeting high coverage (e.g., $\sim$95\%).
\end{enumerate}

\noindent The output is a \emph{Query-Cluster Database} $\mathcal{D}_{qc}$ containing: (i) a fitted projection model, (ii) cluster centroid locations, and (iii) a mapping from cluster ID to recommended retrieval depth (Section~\ref{sec:methodology} details construction). This shifts expensive work---LLM-based relevance labeling and hyperparameter tuning---offline.

\subsubsection{Online: Fast Cluster Assignment}
\label{sec:online}

At inference time, QCAR operates as a thin decision layer preceding standard RAG (Figure~\ref{fig:QCAR}):
\begin{enumerate}
    \item \textbf{Lexical check.} Scan the query for scope markers. If found, set $k^* = k_{\max}$ and proceed to retrieval.
    
    \item \textbf{Cluster assignment.} Otherwise, embed the query, project into the learned space, and assign to the nearest cluster centroid. Look up the corresponding $k^*$ from $\mathcal{D}_{qc}$.
    
    \item \textbf{Retrieval and generation.} Execute standard RAG with the adaptive $k^*$.
\end{enumerate}
\noindent The overhead is minimal: lexical matching (<1ms), one embedding call (30--50ms, already required for retrieval), projection transform ($\sim$1ms), nearest-neighbor search over $K$ centroids (<1ms for $K \leq 100$), and $O(1)$ cluster lookup.

Query distributions drift as user needs evolve. QCAR accommodates this through periodic reconstruction of $\mathcal{D}_{qc}$ or triggered updates when monitoring detects distribution shift. Between updates, the mapping remains static, ensuring consistent behavior.

\subsection{Relationship to Post-Retrieval Methods}
\label{sec:comparison}

QCAR and post-retrieval methods such as CAR~\cite{xu2025clusterbasedadaptiveretrievaldynamic} address complementary failure modes.

\textbf{Complementary strengths.} QCAR excels at \emph{preventing} unnecessary retrieval: simple queries receive small $k^*$ values, saving tokens and latency without sacrificing accuracy. CAR excels at \emph{refining} a retrieved set: given an initial pool, it identifies natural breakpoints in the similarity curve to trim noise.

\textbf{Hybrid designs.} A natural extension combines both: use QCAR to set an initial retrieval budget (a ``prior''), then apply CAR-style cutoff detection on the retrieved set (a ``posterior'' correction). This hybrid inherits QCAR's efficiency gains on easy queries while retaining CAR's adaptive trimming on ambiguous cases. We focus on QCAR alone but note the combination as promising future work.

\subsection{Scope and Assumptions}
\label{sec:scope}

QCAR is designed for \textbf{bounded-domain} RAG systems with stable, recurring query distributions and available historical queries. Documents are drawn from coherent domains (case files, medical records, enterprise wikis) rather than the open web. This characterizes most production RAG deployments: enterprise search, customer support, legal research, clinical decision support, and financial analysis.

\textbf{Domain portability.} While our experiments focus on personal injury law, the methodology is domain-agnostic. Adapting to a new vertical requires: (i) a historical query log, (ii) a relevance labeling mechanism (LLM-as-a-Judge or human annotation), and (iii) optionally, domain-informed lexical patterns. The clustering and depth-estimation machinery transfers directly.

\textbf{Open-domain settings.} General-purpose search engines represent a distinct setting where query diversity is unbounded and corpus characteristics vary unpredictably. QCAR can still contribute by handling recognized query types via learned clusters while deferring novel intents to default $k$ or post-retrieval methods. However, the primary value proposition is in bounded-domain cases.

\section{Methodology}
\label{sec:methodology}

This section details the construction of the Query-Cluster Database $\mathcal{D}_{qc}$ (Sections~\ref{sec:clustering}--\ref{sec:depth-estimation}) and the online inference procedure (Section~\ref{sec:online-inference}).

\subsection{Query Embedding and Clustering}
\label{sec:clustering}

While document clustering is common in post-retrieval reranking, semantic clustering of user queries for adaptive retrieval remains relatively underexplored. Our goal is to group queries with similar search intents to enable cluster-specific retrieval depth optimization.

\subsubsection{Dimensionality Reduction}

We embed historical queries using \texttt{text-embedding-ada-002}~\cite{openai2022embedding}, producing vectors of dimension $N = 1536$. To mitigate the curse of dimensionality~\cite{altman2018curse}, we evaluated two reduction techniques:

\textbf{Principal Component Analysis (PCA)}~\cite{wold1987pca}: A linear technique focused on variance maximization. Cumulative variance analysis showed that 186 components explain 80\% of variance, while 560 components reach 95\%.

\textbf{Uniform Manifold Approximation and Projection (UMAP)}~\cite{mcinnes2018umap, allaoui2020umapstudy}: A non-linear, graph-based approach that preserves local manifold structures. We performed a hyperparameter sweep across \texttt{n\_neighbors}, \texttt{min\_dist}, and \texttt{n\_components} using cosine metric.

\subsubsection{Clustering Algorithms}
We evaluated three clustering methodologies:

\textbf{$K$-means}~\cite{lloyd1982kmeans}: A centroid-based algorithm that minimizes squared Euclidean distances. We apply $L_2$ normalization to ensure that minimizing Euclidean distance is equivalent to maximizing cosine similarity.

\textbf{Agglomerative Clustering with Ward Linkage}~\cite{ward1963hierarchical}: A bottom-up approach that iteratively merges clusters to minimize intra-cluster variance.

\textbf{HDBSCAN}~\cite{campello2013hdbscan}: A density-based method that identifies clusters of varying shapes. Following previous literature~\cite{degroot2022bertopic}, HDBSCAN left approximately one-third of queries unclassified as "noise" despite extensive tuning. We therefore used HDBSCAN to identify the natural number of clusters and stable centroids, then initialized $K$-means to assign all queries to the nearest cluster.

\subsubsection{Evaluation Metrics}

To assess clustering quality, we employ quantitative metrics and LLM-assisted qualitative analysis:

\textbf{Silhouette Score (SS)}~\cite{Shahapure2020cluster}: Measures cluster cohesion and separation on scale $[-1, 1]$. Scores near 1 indicate well-defined clusters; scores near 0 suggest overlap.

\textbf{Davies-Bouldin Index (DBI)}~\cite{davies1979cluster}: Calculates the ratio of intra-cluster dispersion to inter-cluster separation. Lower DBI indicates more distinct clusters.

PCA-paired clustering methods consistently produced SS $\approx 0$ and DBI $> 2$. UMAP was selected for its superior ability to preserve local manifold structures in high-dimensional query embeddings.

Since mathematical metrics often miss semantic nuances, we used \texttt{Gemini 3 Pro}~\cite{google2025gemini} for qualitative analysis. 




\textbf{Final configuration}: UMAP with $N = 2$ achieved the best numerical results (SS $= 0.6$, DBI $= 0.46$), but qualitative analysis revealed it collapses distinct legal topics due to extreme information loss. Balancing mathematical cohesion with semantic granularity, we selected UMAP with \texttt{n\_components}$= 190$, \texttt{n\_neighbors}$= 200$, \texttt{min\_dist}$= 0.0$, and $K = 80$ clusters (SS $= 0.369$, DBI $= 1.019$, high LLM score).
\subsection{Relevance Labeling}
\label{sec:labeling}

We generate ground-truth labels to quantify retrieval performance within clusters by randomly sampling 21 queries from each cluster. Using LLM-as-a-Judge~\cite{zheng2023judgingllmasajudgemtbenchchatbot, liu2023geval}, we employ Claude Sonnet 4.5~\cite{anthropic2025claudesonnet45} with few-shot prompting to label each page (approximately 190 pages per case). 
Each document receives one of three labels with respect to the query: \textbf{0} (not relevant), \textbf{1} (partially relevant), or \textbf{2} (full relevant).

This discrete three-level scheme enables consistent NDCG computation while capturing varying degrees of relevance. It reduces computational complexity and facilitates cluster-level analysis while preserving the ability to differentiate between tangential and central documents. While LLM-based labeling is computationally expensive, this is strictly an offline, one-time cost incurred only during Query-Cluster Database construction.
\subsection{Retrieval Depth Estimation}
\label{sec:depth-estimation}
With the ground-truth labels established, we can now address the core challenge of question complexity. To quantify the difficulty of a specific query within its semantic cluster, we propose using \textbf{Normalized Discounted Cumulative Gain (NDCG)}, a well-established evaluation metric for assessing the quality of ranked retrieval results in information retrieval (IR) systems.
\subsubsection{Why NDCG}
NDCG's widespread adoption stems from its ability to model graded relevance, discount lower-ranked results to reflect realistic user behavior, and normalize scores to enable fair comparison across queries and systems of varying difficulty \cite{jarvelin2002, wang2013, ir_reliability2007}. Consequently, NDCG has become a standard benchmark in major IR evaluation campaigns and learning-to-rank research \cite{valizadegan2009}. 

However, computing NDCG requires constructing an ideal (oracle) ranking for each query, which introduces substantial computational overhead when applied to large-scale query collections. Given the scale of our workload, directly evaluating or optimizing full NDCG rankings is impractical. To address this challenge, we adopt a coarse-grained relevance modeling strategy that partitions retrieved documents into three relevance buckets, as described earlier. This abstraction yields a scalable evaluation framework while preserving the essential distinctions between satisfactory, necessary, and redundant information, enabling efficient and principled analysis of retrieval performance across arbitrary queries while remaining aligned with the core assumptions underlying NDCG-based evaluation.
\subsubsection{NDCG Formulation}
For a query $q$ and its retrieved ranking of documents $\{d_1, d_2, \ldots, d_k, \ldots, d_n\}$, the Discounted Cumulative Gain (DCG) is defined as:
\begin{equation}
\text{DCG}_k = \sum_{i=1}^{k} \frac{2^{\text{rel}_i} - 1}{\log_2(i + 1)}
\end{equation}
where $\text{rel}_i$ is the graded relevance of document $d_i$ with respect to query $q$. The ideal DCG (IDCG) is computed by sorting all relevant documents ($n$) in the optimal order:
\begin{equation}
\text{IDCG}_n = \sum_{i=1}^{n} \frac{2^{\text{rel}_i} - 1}{\log_2(i + 1)}
\end{equation}

The Normalized DCG is then:
\begin{equation}
\text{NDCG}_k = \frac{\text{DCG}_k}{\text{IDCG}_n}, \quad 0 \leq \text{NDCG}_k \leq 1
\end{equation}

A value of 1 indicates perfect retrieval, while lower values indicate missing or poorly ranked relevant documents. Using the LLM relevance labels:
\begin{equation}
\text{NDCG}_k(q) = \frac{\sum_{i=1}^{k} \frac{2^{\text{LLM}(d_i, q)} - 1}{\log_2(i+1)}}{\text{IDCG}_n(q)}
\end{equation}

\begin{figure}[!ht]
    \begin{subfigure}{0.23\textwidth}
    \centering
        \includegraphics[width=\linewidth]{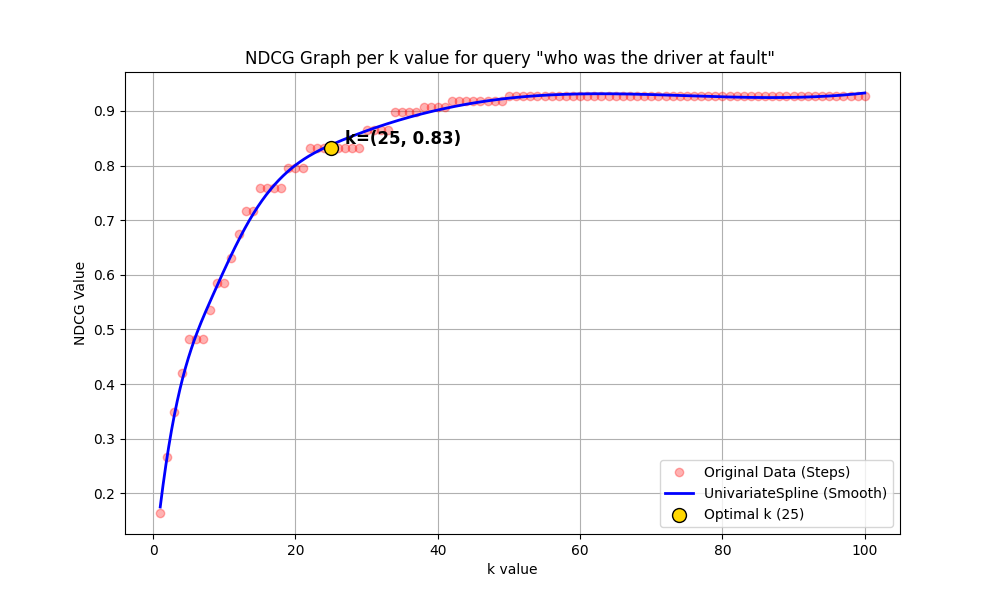}
    \caption{Simple Query NDCG}
    \label{fig:Cluster22-ndcg1}
    \end{subfigure}
    \hfill
    \begin{subfigure}{0.23\textwidth}
    \centering
       \includegraphics[width=\linewidth]{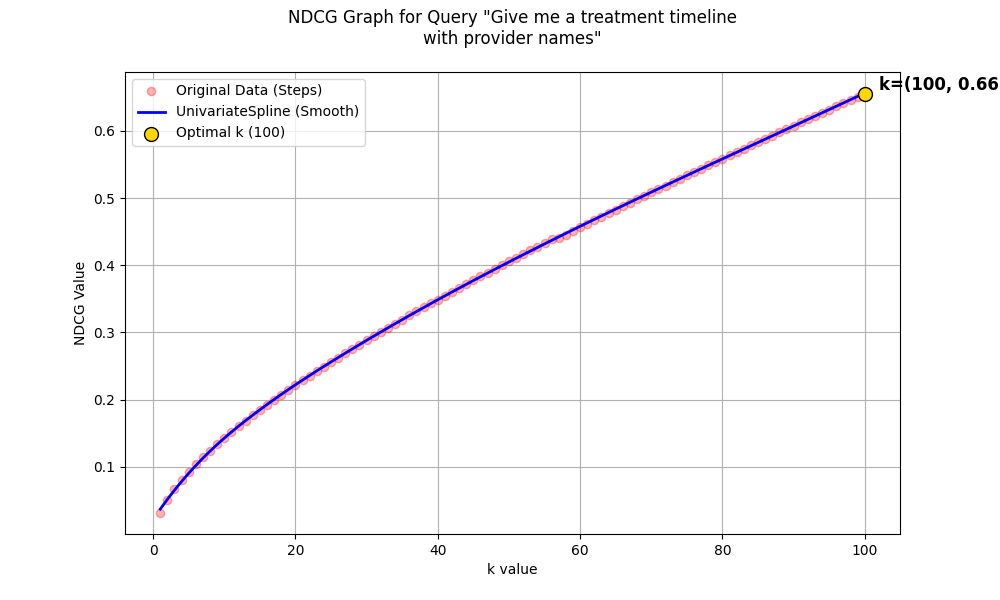}
    \caption{Hard Query NDCG}
    \label{fig:Cluster6-ndcg1}
    \end{subfigure}
    \caption{Samples of NDCG graphs with varying retrieval depth k}
    \Description{Left picture contains NDCG curve for a simple query where it clearly shows an elbow point. Right picture contains NDCG curve for a hard query with a steady growth}
    \label{fig:ndcg-curves}
\end{figure}

Intuitively, NDCG serves as a lens for question complexity: questions requiring broad, diffuse, or highly contextual retrieval tend to produce lower NDCG under default retrieval settings, while simpler questions yield steeper relevance curves with high NDCG (Figure~\ref{fig:ndcg-curves}).

\subsubsection{Per-Query Elbow Point}
Our approach is as follows: for each query $q$, we execute RAG retrieval with its maximum allowed parameters ($k=100$) and compute the resulting NDCG at varying retrieval depths from $k=1$ to $k=100$ with the ground-truth page relevancy labels.

We define a query-specific recommended $k^*(q)$ as the maximum $k$ such that NDCG gains remain non-negligible (with threshold $\epsilon = 0.001$):
\begin{equation}
k^*(q) = \max\left\{k \mid \text{NDCG}_{k+1}(q) - \text{NDCG}_k(q) \geq \epsilon\right\}
\end{equation}

Figure~\ref{fig:ndcg-curves} illustrates representative NDCG-$k$ curves. Simple queries (e.g., "Who was the driver at fault?") demonstrate rapid convergence with an elbow point at $k = 25$ with NDCG increasing by only 0.1 across the remaining 75 documents. Hard queries (e.g., "List all treatments with dates and providers") exhibit sustained growth up to $k = 100$ without a defined plateau, indicating information is distributed more broadly across the document set.

\subsubsection{Cluster-Level Aggregation}

Let a cluster of questions $C_j = \{q_1, q_2, \ldots, q_{n_j}\}$ have $n_j$ queries. For this cluster, we compute the mean and standard deviation of the query-specific $k^*$ values:
\begin{equation}
\mu_j = \frac{1}{n_j} \sum_{i=1}^{n_j} k^*(q_i), \quad \sigma_j = \sqrt{\frac{1}{n_j} \sum_{i=1}^{n_j} \left(k^*(q_i) - \mu_j\right)^2}
\end{equation}

To ensure coverage of approximately 95\% of relevant documents within the cluster, we define the cluster-level recommended $k^*$ as:
\begin{equation}
k^*_{\text{cluster}} = \mu_j + 2\sigma_j
\end{equation}

At query time, an incoming query $q^*$ assigned to cluster $C_j$ will use $k^*_{\text{cluster}}$ as the RAG top-$k$ parameter, providing a retrieval depth sufficient to capture most relevant documents while avoiding unnecessary over-retrieval.

Importantly, this method scales efficiently: cluster membership depends only on the dimension of the clustering space and number of clusters, not on the number of documents per case or query, giving constant-time complexity at query time.

\subsection{Online Inference}
\label{sec:online-inference}

At inference time, the QCAR framework operates via a decoupled architecture: the heavy-compute offline phase produces $\mathcal{D}_{qc}$, while a lightweight online phase handles query routing. The inference procedure follows these steps (Algorithm~\ref{alg:qcar}):

\SetKwSty{textbf} 
\SetArgSty{textnormal}
\begin{algorithm}[t]
\small
\SetKwInOut{Input}{Input}\SetKwInOut{Output}{Output}
\SetKwFunction{UMAP}{UMAP\_Project}
\SetKwFunction{Dist}{EuclideanDist}
\SetKwFunction{Match}{LexicalMatch}
\Input{User Query $q$, Pre-computed Database $\mathcal{D}_{qc}$, Max Retrieval Depth $k_{max}$}
\Output{Adaptive Retrieval Parameter $k^*$, Generated Response $R$}
\BlankLine
\tcp{Stage 1: Lexical Heuristics}
\If{\Match{$q$}}{
    $k^* \gets k_{max}$\;
}
\Else{
    \tcp{Stage 2: QCAR Semantic Mapping}
    $v_q \gets \text{Embedding}(q)$\;
    $\hat{v}_q \gets$ \UMAP{$v_q, \mathcal{D}_{qc}.\text{model}$}\;
    $c^* \gets \arg\min_{c \in \mathcal{D}_{qc}.\text{centroids}} \text{Dist}(\hat{v}_q, c)$ \tcp*[r]{Nearest Cluster}
    $k^* \gets \mathcal{D}_{qc}.\text{map}(c^*)$\;
}
\BlankLine
\tcp{Stage 3: RAG Execution}
$\mathcal{C} \gets \text{Retrieve}(q, k^*)$ \tcp*[r]{Context Documents}
$R \gets \text{LLM-Generate}(q, \mathcal{C})$\;
\Return $R, k^*$\;
\caption{QCAR Adaptive Retrieval Inference}
\label{alg:qcar}
\end{algorithm}

\textbf{Step 1: Lexical Heuristics.} The system first checks for predefined lexical patterns. Queries exhibiting these lexical or structural signals are assigned $k^* = k_{\max}$ and proceed directly to retrieval. Approximately 80\% of queries do not fall into these predefined categories, highlighting the necessity of the clustering approach for the majority of queries.

\textbf{Step 2: QCAR Semantic Mapping.} If no lexical match occurs, the QCAR model performs a low-latency semantic mapping in real-time while accessing cluster and $k^*$ mappings via the precomputed $\mathcal{D}_{qc}$. This process involves a single embedding call; UMAP transformation using the fitted projection model, nearest-neighbor centroid calculation to determine cluster assignment and lookup of the corresponding $k^*_{\text{cluster}}$.

\textbf{Step 3: RAG Execution.} RAG retrieves exactly $k^*$ documents, which then undergo the standard downstream processing within the existing RAG infrastructure.

Query-Cluster Database $\mathcal{D}_{qc}$ contains the following information: (i) the fitted UMAP projection model, (ii) cluster centroid locations, and (iii) the cluster-to-$k^*$ mapping table. These components are accessed by the QCAR model as part of the RAG framework at inference time.
\section{Experiments}
\label{sec:experiments}

\subsection{Data Preparation}
The initial dataset comprised 20,000 raw user queries. Following a rigorous cleaning process which involved removing duplicates, filtering for Personally Identifiable Information (PII), etc, we obtained 6,820 questions in total.
We also conducted a lexical analysis to categorize the questions, identifying four common lexical patterns relevant to the legal domain. Approximately 80\% of the questions did not fall into any of these predefined categories. This gap highlights the limitations of rule-based lexical patterns and underscores the necessity of utilizing clustering methods to better distinguish user intent.

\subsection{Clustering Results}
UMAP with k-means ($\texttt{n\_components} = 2$, $\texttt{n\_neighbors}=200$, $\texttt{min\_dist}=0.0$ and $k=60$) provided the best numerical results, where $SS = 0.6$ and $DBI = 0.46$. However, qualitative analysis revealed that while $N=2$ creates tight visual clusters, it collapses distinct legal topics into single groups due to extreme information loss.

To balance mathematical cohesion with semantic granularity, we extended our analysis to higher UMAP dimensions. By synthesizing the results of the mathematical evaluation and the qualitative assessment from the LLM-as-a-Judge, we determined that the optimal balance between semantic preservation and cluster distinctness was achieved using UMAP with ($\texttt{n\_components} = 190, \texttt{n\_neighbors}=200, \texttt{min\_dist}=0.0$ and $k=80$) clusters.

\subsection{NDCG Results}

Having identified optimal clusters, we perform NDCG-based optimization to determine the optimal retrieval depth $k^*$ for each cluster.

\subsubsection{Cluster-Specific Depth Analysis} 
We evaluate our proposed adaptive top-$k$ strategy against a static baseline of $k=50$, which represents a highly optimized production configuration of our live system at the time of this study. To provide a rigorous evaluation of the adaptive mechanism, we also benchmark performance against other static settings. Due to infrastructure constraints and LLM context window limits, the adaptive selection $k^*$ is capped at a maximum threshold of 100. Table~\ref{tab:f1 changes} showcases the performance gains and token trade-offs of our system, partitioned by query behavior into token-decreasing (simple) and token-increasing (complex) subsets.

For simple, fact-retrieval queries (e.g., plaintiff's date of birth), the system achieves stable NDCG at lower retrieval depths. Against the $k=50$ baseline, the adaptive approach yields a $35.95\%$ reduction in token consumption while simultaneously improving the $F_1$ score by $+0.37$. Complex queries (e.g., list all treatments and procedures with dates and providers), require broader document coverage, with the optimal $k^*$ frequently reaching our 100-document upper bound. Compared to the $k=50$ baseline, this expanded retrieval improves the $F_1$ score by $0.37$ at the cost of an $85.95\%$ increase in token consumption. This positive correlation between retrieval depth and $F_1$ underscores the necessity of extensive coverage for complex information needs, proving that a static retrieval depth is fundamentally sub-optimal.
\begin{table}[t] 
\centering
\small
\caption{Performance of adaptive $k$ split by query behavior. Both token-decreasing and token-increasing query subsets yield improvements in $F_1$ score compared to the static top-$k$ baseline.}
\label{tab:f1 changes}
\begin{tabular*}{\columnwidth}{@{\extracolsep{\fill}} l cccc @{}}
\toprule
\textbf{Against Baseline} & \multicolumn{2}{c}{\textbf{\begin{tabular}[b]{@{}c@{}}Token-Increasing\\ Queries\end{tabular}}} & \multicolumn{2}{c}{\textbf{\begin{tabular}[b]{@{}c@{}}Token-Decreasing\\ Queries\end{tabular}}} \\ 
\cmidrule(lr){2-3} \cmidrule(lr){4-5}
\textbf{top\_k}           & \textbf{$\Delta$ $F_1$} & \textbf{$\Delta$ Tok.}          & \textbf{$\Delta$ $F_1$} & \textbf{$\Delta$ Tok.} \\ \midrule
top\_k = 25               & \textbf{+0.94}       & +244.88\%                       & \textbf{+0.25}       & -27.68\%               \\
top\_k = 50               & \textbf{+0.37}       & +85.95\%                        & \textbf{+0.37}       & -35.95\%               \\
top\_k = 75               & \textbf{+0.15}       & +32.24\%                        & \textbf{+0.40}       & -38.42\%               \\
top\_k = 100              & ---                  & ---                             & \textbf{+0.46}       & -40.77\%               \\
\bottomrule
\end{tabular*}
\end{table}

\subsection{Complexity-Aware Dynamic Retrieval Performance}
\subsubsection{Overall Metric Comparison}

\begin{figure}[t]
    \centering
    \begin{subfigure}[t]{0.23\textwidth}
        \centering
        \includegraphics[width=\linewidth]{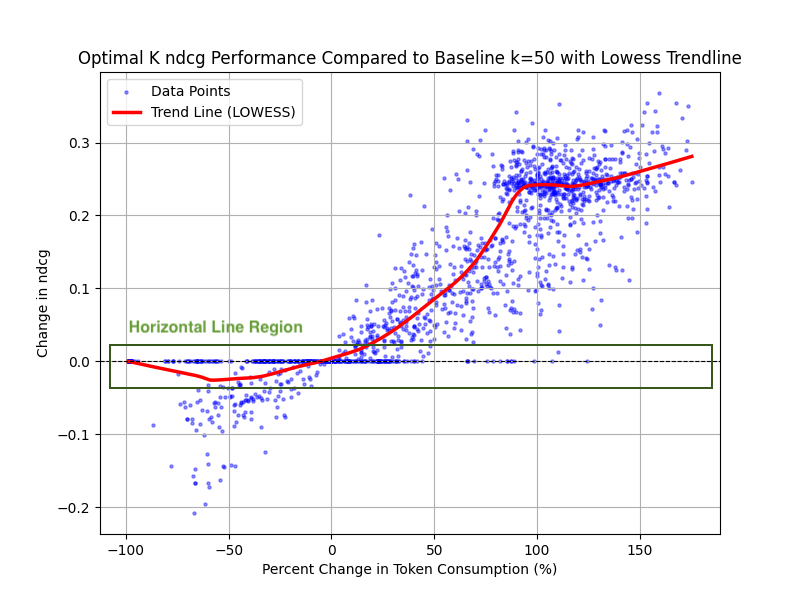}
        \caption{Adaptive $k$ - NDCG}
        \label{fig:adaptivendcg}
    \end{subfigure}
    \hfill 
    \begin{subfigure}[t]{0.23\textwidth}
        \centering
        \includegraphics[width=\linewidth]{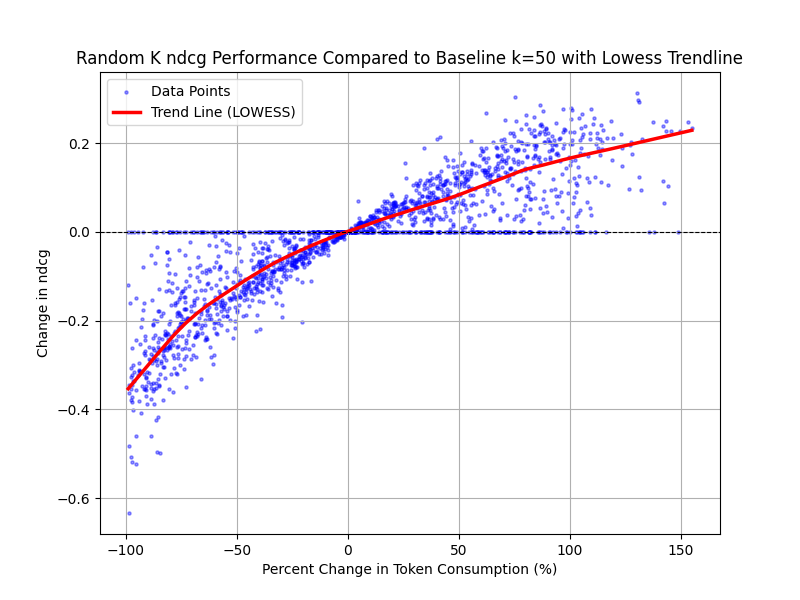}
        \caption{Random $k$ - NDCG}
        \label{fig:randomndcg}
    \end{subfigure}
    \begin{subfigure}[t]{0.23\textwidth}
        \centering
        \includegraphics[width=\linewidth]{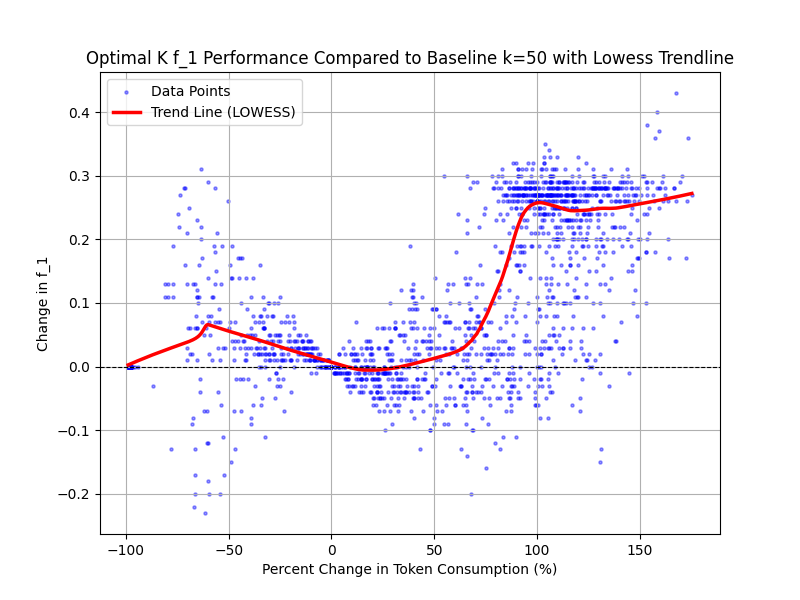}
    \caption{Adaptive $k$ - F-1 Score}
    \label{fig:adaptivef1}
    \end{subfigure}
    \hfill 
    \begin{subfigure}[t]{0.23\textwidth}
        \centering
       \includegraphics[width=\linewidth]{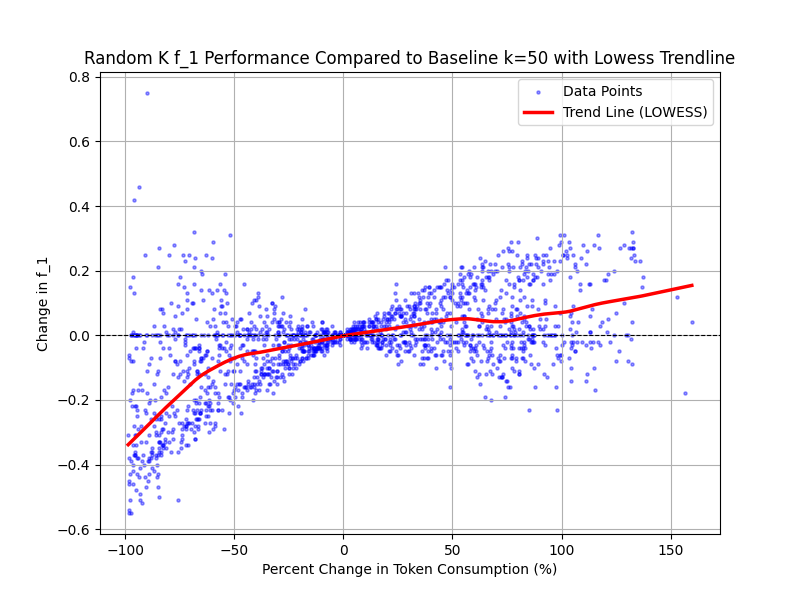}
    \caption{Random $k$ - F-1 Score }
    \label{fig:randomf1}
    \end{subfigure}       
    \caption{Performance vs. baseline $k=50$: adaptive strategy (left) maintains accuracy with fewer tokens; random strategy (right) degrades immediately.}
    \Description{Four scatter plots comparing the changes in retrieval performance against the changes in token consumption relative to a baseline of $k=50$. The top row reports NDCG performance while the bottom row reports $F_1$ score; the left column uses the proposed adaptive retrieval depth and the right column uses a randomly selected retrieval depth. For adaptive top-k, performance remains approximately unchanged when token consumption decreases. In contrast, random reductions in top-k results in performance degradation as token consumption decreases. LOWESS trend lines show the overall relationship in each plot.
}
    \label{fig:metricperformance}
\end{figure}

Figure~\ref{fig:metricperformance} compares the adaptive $k$ strategy against a randomly selected $k$ baseline across NDCG, and $F_1$ metrics. LOWESS trendlines highlight overall behavior. The results demonstrate that adaptive retrieval optimizes token consumption without sacrificing retrieval quality.

\textbf{Token Efficiency:} The horizontal region ($y \approx 0$) in NDCG plots (Fig.~\ref{fig:adaptivendcg}) shows that adaptive $k$ maintains baseline performance despite token reductions up to 100\% (relative to $k=50$). This high density of points at $y=0$ in the negative $x$-quadrant confirms that dynamic selection effectively identifies and discards redundant documents without harming retrieval accuracy.

In the lower-left quadrant (both token consumption and metrics $<0$), the random $k$ baseline exhibits steep, linear decline in NDCG (Figs.~\ref{fig:randomndcg}), while the adaptive strategy maintains a robust plateau. Random context reduction frequently omits essential documents, whereas QCAR preserves relevant content while truncating simple queries.

\textbf{$F_1$ Score and Generation Quality:} The $F_1$ trends confirm that stochastic document reduction diminishes retrieval quality (Fig.~\ref{fig:randomf1}). In contrast, the stability of adaptive $k$ (Fig.~\ref{fig:adaptivef1}) demonstrates that QCAR preserves essential content for RAG generation while maximizing cost efficiency.

\subsubsection{A/B Test Results}
An A/B test was executed on a uniform sample of live queries under full serving conditions. The experiment compared the default static \(k\) configuration against a dynamic \(k\) variant, using an ensemble of language models. Changes in token consumption and model quality were measured.
Launch approval was granted with the following Tier-1 metrics:
\begin{itemize}
\item Token consumption increased by \(51.19\%\) (\(p = 2.87 \times 10^{-134}\)).
\item $F_1$ score improved by \(36.64\%\) (\(p = 1.14 \times 10^{-34}\)).
\end{itemize}
\subsubsection{End-to-End Analysis}  
To evaluate the downstream impact of the dynamic $k$ retrieval strategy, we sampled user queries from both the simple and complex clusters, evaluating the generated responses via an LLM-as-a-Judge framework. 

The dynamic mechanism reduces token consumption up to 23\% while maintaining conceptual parity with the baseline for simple queries. It also improves factual recall by $9\%$ to $17\%$ for complex queries. This demonstrates that the adaptive approach successfully captures distributed evidence for sophisticated inquiries without incurring unnecessary computational overhead on simpler tasks.

\section{Conclusion}
\label{sec:conclusion}
This paper introduced QCAR (Query-Clustering Adaptive Retrieval), a pre-retrieval framework addressing the mismatch between static top-k retrieval and queries spanning orders of magnitude in evidence requirements. Through offline UMAP-based clustering and NDCG-driven depth estimation, QCAR enables constant-time adaptive retrieval via lexical override for explicitly scoped queries and cluster-based depth selection otherwise. A/B testing demonstrated dual benefits: up to 23\% token reduction on simple queries without accuracy loss while expanding context for complex synthesis, achieving 36\% overall $F_1$ improvement and up to 17\% more evidence coverage on comprehensive queries. Though validated in personal injury law, the modular architecture—offline cluster construction with lightweight online inference—transfers directly to any bounded-domain RAG system (healthcare, finance, enterprise search) where historical queries reveal recurring patterns, offering a practical complement to post-retrieval methods.

\section*{GenAI Usage Disclosure}
\label{sec:disclosure}
We acknowledge the use of Generative AI (GenAI) tools in the
preparation of this work. GenAI tools were only used to improve
spelling and grammar of certain passages. All uses of GenAI tools
complied with the ACM Authorship Policy on GenAI usage.
\bibliographystyle{ACM-Reference-Format}
\balance
\bibliography{bibfile}
\end{document}